\documentclass[longauth]{aa} 
\usepackage{color}
\usepackage{comment}
\usepackage{natbib}
\usepackage{url}
\usepackage{hyperref}
\usepackage{subfigure}
\usepackage{xspace}
\usepackage[percent]{overpic}
\usepackage{subfloat}

\newcommand{\ixpe}{{\it IXPE}\xspace}
\newcommand{\artxc}{{ART-XC}\xspace}
\newcommand{\hxmt}{{\it HXMT}\xspace}

\newcommand{\src}{EXO~2030+375}
\graphicspath{{./}{figures/}}

\begin{document}

\title{A polarimetrically oriented X-ray stare at the accreting pulsar \src}

\titlerunning{X-ray polarization of X-ray pulsar \src}
\authorrunning{C. Malacaria et al.}


\author{Christian~Malacaria \inst{\ref{in:ISSI}}
\and Jeremy~Heyl \inst{\ref{in:UBC}}
\and Victor~Doroshenko \inst{\ref{in:Tub}}
\and Sergey~S.~Tsygankov \inst{\ref{in:UTU}}
\and Juri Poutanen \inst{\ref{in:UTU}}
\and Sofia~V.~Forsblom \inst{\ref{in:UTU}}
\and Fiamma~Capitanio \inst{\ref{in:INAF-IAPS}} 
\and Alessandro~Di~Marco \inst{\ref{in:INAF-IAPS}}
\and Yujia~Du \inst{\ref{in:Tub}}
\and Lorenzo~Ducci \inst{\ref{in:Tub},\ref{in:ISDC}}
\and Fabio~La~Monaca \inst{\ref{in:INAF-IAPS}} 
\and Alexander~A.~Lutovinov \inst{\ref{in:IKI}}
\and Herman~L.~Marshall \inst{\ref{in:MIT}}
\and Ilya~A. Mereminskiy \inst{\ref{in:IKI}}
\and Sergey~V.~Molkov \inst{\ref{in:IKI}}
\and Alexander A. Mushtukov\inst{\ref{in:Oxford}}
\and Mason~Ng \inst{\ref{in:MIT}}
\and Pierre-Olivier~Petrucci \inst{\ref{in:Grenoble}} 
\and Andrea~Santangelo \inst{\ref{in:Tub}} 
\and Andrey~E.~Shtykovsky \inst{\ref{in:IKI}} 
\and Valery~F.~Suleimanov \inst{\ref{in:Tub}} 
\and Iv\'an~Agudo \inst{\ref{in:CSIC-IAA}}
\and Lucio~A.~Antonelli \inst{\ref{in:INAF-OAR},\ref{in:ASI-SSDC}} 
\and Matteo~Bachetti \inst{\ref{in:INAF-OAC}} 
\and Luca~Baldini  \inst{\ref{in:INFN-PI},      \ref{in:UniPI}} 
\and Wayne~H.~Baumgartner  \inst{\ref{in:NASA-MSFC}}
\and Ronaldo~Bellazzini  \inst{\ref{in:INFN-PI}} 
\and Stefano~Bianchi \inst{\ref{in:UniRoma3}}  
\and Stephen~D.~Bongiorno \inst{\ref{in:NASA-MSFC}} 
\and Raffaella~Bonino  \inst{\ref{in:INFN-TO},\ref{in:UniTO}}
\and Alessandro~Brez  \inst{\ref{in:INFN-PI}} 
\and Niccol\`{o}~Bucciantini 
\inst{\ref{in:INAF-Arcetri},\ref{in:UniFI},\ref{in:INFN-FI}} 
\and Simone~Castellano \inst{\ref{in:INFN-PI}} 
\and Elisabetta~Cavazzuti \inst{\ref{in:ASI}} 
\and Chien-Ting~Chen \inst{\ref{in:USRA-MSFC}}
\and Stefano~Ciprini \inst{\ref{in:INFN-Roma2},\ref{in:ASI-SSDC}}
\and Enrico~Costa \inst{\ref{in:INAF-IAPS}} 
\and Alessandra~De~Rosa \inst{\ref{in:INAF-IAPS}} 
\and Ettore~Del~Monte \inst{\ref{in:INAF-IAPS}} 
\and Laura~Di~Gesu \inst{\ref{in:ASI}} 
\and Niccol\`{o}~Di~Lalla \inst{\ref{in:Stanford}}
\and Immacolata~Donnarumma \inst{\ref{in:ASI}}
\and Michal~Dov\v{c}iak \inst{\ref{in:CAS-ASU}}
\and Steven~R.~Ehlert \inst{\ref{in:NASA-MSFC}}  
\and Teruaki~Enoto \inst{\ref{in:RIKEN}}
\and Yuri~Evangelista \inst{\ref{in:INAF-IAPS}}
\and Sergio~Fabiani \inst{\ref{in:INAF-IAPS}}
\and Riccardo~Ferrazzoli \inst{\ref{in:INAF-IAPS}} 
\and Javier~A.~Garcia \inst{\ref{in:Caltech}}
\and Shuichi~Gunji \inst{\ref{in:Yamagata}} 
\and Kiyoshi~Hayashida \inst{\ref{in:Osaka}}\thanks{Deceased.}  
\and Wataru~Iwakiri \inst{\ref{in:Chiba}} 
\and Svetlana~G.~Jorstad \inst{\ref{in:BU},\ref{in:SPBU}} 
\and Philip~Kaaret \inst{\ref{in:NASA-MSFC}}  
\and Vladimir~Karas \inst{\ref{in:CAS-ASU}}
\and Fabian~Kislat \inst{\ref{in:UNH}} 
\and Takao~Kitaguchi  \inst{\ref{in:RIKEN}} 
\and Jeffery~J.~Kolodziejczak \inst{\ref{in:NASA-MSFC}} 
\and Henric~Krawczynski  \inst{\ref{in:WUStL}}
\and Luca~Latronico  \inst{\ref{in:INFN-TO}} 
\and Ioannis~Liodakis \inst{\ref{in:FINCA}}
\and Simone~Maldera \inst{\ref{in:INFN-TO}}  
\and Alberto~Manfreda \inst{\ref{INFN-NA}}
\and Fr\'{e}d\'{e}ric~Marin \inst{\ref{in:Strasbourg}} 
\and Andrea~Marinucci \inst{\ref{in:ASI}} 
\and Alan~P.~Marscher \inst{\ref{in:BU}} 
\and Francesco~Massaro \inst{\ref{in:INFN-TO},\ref{in:UniTO}} 
\and Giorgio~Matt  \inst{\ref{in:UniRoma3}}  
\and Ikuyuki~Mitsuishi \inst{\ref{in:Nagoya}} 
\and Tsunefumi~Mizuno \inst{\ref{in:Hiroshima}} 
\and Fabio~Muleri \inst{\ref{in:INAF-IAPS}} 
\and Michela~Negro \inst{\ref{in:UMBC},\ref{in:NASA-GSFC},\ref{in:CRESST}} 
\and Chi-Yung~Ng \inst{\ref{in:HKU}}
\and Stephen~L.~O'Dell \inst{\ref{in:NASA-MSFC}}  
\and Nicola~Omodei \inst{\ref{in:Stanford}}
\and Chiara~Oppedisano \inst{\ref{in:INFN-TO}}  
\and Alessandro~Papitto \inst{\ref{in:INAF-OAR}}
\and George~G.~Pavlov \inst{\ref{in:PSU}}
\and Abel~L.~Peirson \inst{\ref{in:Stanford}}
\and Matteo~Perri \inst{\ref{in:ASI-SSDC},\ref{in:INAF-OAR}}
\and Melissa~Pesce-Rollins \inst{\ref{in:INFN-PI}} 
\and Maura~Pilia \inst{\ref{in:INAF-OAC}} 
\and Andrea~Possenti \inst{\ref{in:INAF-OAC}} 
\and Simonetta~Puccetti \inst{\ref{in:ASI-SSDC}}
\and Brian~D.~Ramsey \inst{\ref{in:NASA-MSFC}} 
\and John~Rankin \inst{\ref{in:INAF-IAPS}} 
\and Ajay~Ratheesh \inst{\ref{in:INAF-IAPS}} 
\and Oliver~J.~Roberts \inst{\ref{in:USRA-MSFC}}
\and Roger~W.~Romani \inst{\ref{in:Stanford}}
\and Carmelo~Sgr\`o \inst{\ref{in:INFN-PI}}  
\and Patrick~Slane \inst{\ref{in:CfA}}  
\and Paolo~Soffitta \inst{\ref{in:INAF-IAPS}} 
\and Gloria~Spandre \inst{\ref{in:INFN-PI}} 
\and Douglas~A.~Swartz \inst{\ref{in:USRA-MSFC}}
\and Toru~Tamagawa \inst{\ref{in:RIKEN}}
\and Fabrizio~Tavecchio \inst{\ref{in:INAF-OAB}}
\and Roberto~Taverna \inst{\ref{in:UniPD}}  
\and Yuzuru~Tawara \inst{\ref{in:Nagoya}}
\and Allyn~F.~Tennant \inst{\ref{in:NASA-MSFC}}
\and Nicholas~E.~Thomas \inst{\ref{in:NASA-MSFC}}  
\and Francesco~Tombesi  \inst{\ref{in:UniRoma2},\ref{in:INFN-Roma2},\ref{in:UMd}}
\and Alessio~Trois \inst{\ref{in:INAF-OAC}}
\and Roberto~Turolla \inst{\ref{in:UniPD},\ref{in:MSSL}}
\and Jacco~Vink \inst{\ref{in:Amsterdam}}
\and Martin~C.~Weisskopf \inst{\ref{in:NASA-MSFC}} 
\and Kinwah~Wu \inst{\ref{in:MSSL}}
\and Fei~Xie \inst{\ref{in:GSU},\ref{in:INAF-IAPS}}
\and Silvia~Zane  \inst{\ref{in:MSSL}}
}

\institute{International Space Science Institute, Hallerstrasse 6, 3012 Bern, Switzerland \label{in:ISSI}
\\ \email{cmalacaria.astro@gmail.com}
\and 
University of British Columbia, Vancouver, BC V6T 1Z4, Canada \label{in:UBC}
\and
Institut f\"ur Astronomie und Astrophysik, Universit\"at T\"ubingen, Sand 1, D-72076 T\"ubingen, Germany \label{in:Tub}
\and
Department of Physics and Astronomy, FI-20014 University of Turku,  Finland \label{in:UTU} 
\and  INAF Istituto di Astrofisica e Planetologia Spaziali, Via del Fosso del Cavaliere 100, 00133 Roma, Italy \label{in:INAF-IAPS}
\and
ISDC Data Center for Astrophysics, Universit\'e de Gen\`eve, 16 chemin d'\'Ecogia, 1290 Versoix, Switzerland \label{in:ISDC}
\and Space Research Institute (IKI) of Russian Academy of Sciences, Prosoyuznaya ul 84/32, 117997 Moscow, Russian Federation \label{in:IKI}
\and 
MIT Kavli Institute for Astrophysics and Space Research, Massachusetts Institute of Technology, 77 Massachusetts Avenue, Cambridge, MA 02139, USA \label{in:MIT}
\and Astrophysics, Department of Physics, University of Oxford, Denys Wilkinson Building, Keble Road, Oxford OX1 3RH, UK \label{in:Oxford}
\and 
Universit\'{e} Grenoble Alpes, CNRS, IPAG, 38000 Grenoble, France \label{in:Grenoble}
\and 
Instituto de Astrof\'{i}sicade Andaluc\'{i}a -- CSIC, Glorieta de la Astronom\'{i}a s/n, 18008 Granada, Spain \label{in:CSIC-IAA}
\and 
INAF Osservatorio Astronomico di Roma, Via Frascati 33, 00040 Monte Porzio Catone (RM), Italy \label{in:INAF-OAR}  
\and 
Space Science Data Center, Agenzia Spaziale Italiana, Via del Politecnico snc, 00133 Roma, Italy \label{in:ASI-SSDC}
 \and
INAF Osservatorio Astronomico di Cagliari, Via della Scienza 5, 09047 Selargius (CA), Italy  \label{in:INAF-OAC}
\and 
Istituto Nazionale di Fisica Nucleare, Sezione di Pisa, Largo B. Pontecorvo 3, 56127 Pisa, Italy \label{in:INFN-PI}
\and  
Dipartimento di Fisica, Universit\`{a} di Pisa, Largo B. Pontecorvo 3, 56127 Pisa, Italy \label{in:UniPI} 
\and 
NASA Marshall Space Flight Center, Huntsville, AL 35812, USA \label{in:NASA-MSFC}
\and 
Dipartimento di Matematica e Fisica, Universit\`a degli Studi Roma Tre, via della Vasca Navale 84, 00146 Roma, Italy  \label{in:UniRoma3}
\and  
Istituto Nazionale di Fisica Nucleare, Sezione di Torino, Via Pietro Giuria 1, 10125 Torino, Italy  \label{in:INFN-TO}      
\and  
Dipartimento di Fisica, Universit\`{a} degli Studi di Torino, Via Pietro Giuria 1, 10125 Torino, Italy \label{in:UniTO} 
\and   
INAF Osservatorio Astrofisico di Arcetri, Largo Enrico Fermi 5, 50125 Firenze, Italy 
\label{in:INAF-Arcetri} 
\and  
Dipartimento di Fisica e Astronomia, Universit\`{a} degli Studi di Firenze, Via Sansone 1, 50019 Sesto Fiorentino (FI), Italy \label{in:UniFI} 
\and   
Istituto Nazionale di Fisica Nucleare, Sezione di Firenze, Via Sansone 1, 50019 Sesto Fiorentino (FI), Italy \label{in:INFN-FI}
\and 
Agenzia Spaziale Italiana, Via del Politecnico snc, 00133 Roma, Italy \label{in:ASI}
\and 
Science and Technology Institute, Universities Space Research Association, Huntsville, AL 35805, USA \label{in:USRA-MSFC}
\and 
Istituto Nazionale di Fisica Nucleare, Sezione di Roma ``Tor Vergata'', Via della Ricerca Scientifica 1, 00133 Roma, Italy 
 \label{in:INFN-Roma2}
\and 
Department of Physics and Kavli Institute for Particle Astrophysics and Cosmology, Stanford University, Stanford, California 94305, USA  \label{in:Stanford}
\and 
Astronomical Institute of the Czech Academy of Sciences, Boční II 1401/1, 14100 Praha 4, Czech Republic \label{in:CAS-ASU}
\and 
RIKEN Cluster for Pioneering Research, 2-1 Hirosawa, Wako, Saitama 351-0198, Japan \label{in:RIKEN}
\and 
California Institute of Technology, Pasadena, CA 91125, USA \label{in:Caltech}
\and Yamagata University,1-4-12 Kojirakawa-machi, Yamagata-shi 990-8560, Japan \label{in:Yamagata}
\and 
Osaka University, 1-1 Yamadaoka, Suita, Osaka 565-0871, Japan \label{in:Osaka}
\and 
International Center for Hadron Astrophysics, Chiba University, Chiba 263-8522, Japan \label{in:Chiba}
\and
Institute for Astrophysical Research, Boston University, 725 Commonwealth Avenue, Boston, MA 02215, USA \label{in:BU} 
\and 
Department of Astrophysics, St. Petersburg State University, Universitetsky pr. 28, Petrodvoretz, 198504 St. Petersburg, Russia \label{in:SPBU} 
\and 
Department of Physics and Astronomy and Space Science Center, University of New Hampshire, Durham, NH 03824, USA \label{in:UNH} 
\and 
Physics Department and McDonnell Center for the Space Sciences, Washington University in St. Louis, St. Louis, MO 63130, USA \label{in:WUStL}
\and 
Finnish Centre for Astronomy with ESO,  20014 University of Turku, Finland \label{in:FINCA}
\and 
Istituto Nazionale di Fisica Nucleare, Sezione di Napoli, Strada Comunale Cinthia, 80126 Napoli, Italy \label{INFN-NA}
\and 
Universit\'{e} de Strasbourg, CNRS, Observatoire Astronomique de Strasbourg, UMR 7550, 67000 Strasbourg, France \label{in:Strasbourg}
\and 
Graduate School of Science, Division of Particle and Astrophysical Science, Nagoya University, Furo-cho, Chikusa-ku, Nagoya, Aichi 464-8602, Japan \label{in:Nagoya}
\and 
Hiroshima Astrophysical Science Center, Hiroshima University, 1-3-1 Kagamiyama, Higashi-Hiroshima, Hiroshima 739-8526, Japan \label{in:Hiroshima}
\and  
University of Maryland, Baltimore County, Baltimore, MD 21250, USA \label{in:UMBC}
\and 
NASA Goddard Space Flight Center, Greenbelt, MD 20771, USA  \label{in:NASA-GSFC}
\and 
Center for Research and Exploration in Space Science and Technology, NASA/GSFC, Greenbelt, MD 20771, USA  \label{in:CRESST}
\and 
Department of Physics, University of Hong Kong, Pokfulam, Hong Kong \label{in:HKU}
\and 
Department of Astronomy and Astrophysics, Pennsylvania State University, University Park, PA 16801, USA \label{in:PSU}
\and 
Center for Astrophysics, Harvard \& Smithsonian, 60 Garden St, Cambridge, MA 02138, USA \label{in:CfA} 
\and 
INAF Osservatorio Astronomico di Brera, via E. Bianchi 46, 23807 Merate (LC), Italy \label{in:INAF-OAB}
\and 
Dipartimento di Fisica e Astronomia, Universit\`{a} degli Studi di Padova, Via Marzolo 8, 35131 Padova, Italy \label{in:UniPD}
\and
Dipartimento di Fisica, Universit\`{a} degli Studi di Roma ``Tor Vergata'', Via della Ricerca Scientifica 1, 00133 Roma, Italy \label{in:UniRoma2}
\and
Department of Astronomy, University of Maryland, College Park, Maryland 20742, USA \label{in:UMd}
\and 
Mullard Space Science Laboratory, University College London, Holmbury St Mary, Dorking, Surrey RH5 6NT, UK \label{in:MSSL}
\and 
Anton Pannekoek Institute for Astronomy \& GRAPPA, University of Amsterdam, Science Park 904, 1098 XH Amsterdam, The Netherlands  \label{in:Amsterdam}
\and 
Guangxi Key Laboratory for Relativistic Astrophysics, School of Physical Science and Technology, Guangxi University, Nanning 530004, China \label{in:GSU}
}

\abstract
{Accreting X-ray pulsars (XRPs) are presumed to be ideal targets for polarization measurements, as their high magnetic field strength is expected to polarize the emission up to a polarization degree of $\sim$80\%.
However, such expectations are being challenged by recent observations of XRPs with the Imaging X-ray Polarimeter Explorer (\ixpe).
Here, we report on the results of yet another XRP, namely, \src, observed with \ixpe and contemporarily monitored with {\it Insight-HXMT} and {\it SRG}/ART-XC.
In line with recent results obtained with \ixpe for similar sources, an analysis of the \src\ data returns a low polarization degree of 0\%--3\% 
in the phase-averaged study and a variation in the range of 2\%--7\% in the phase-resolved study.
Using the rotating vector model, we constrained the geometry of the system and obtained a value of $\sim$60\degr \ for the magnetic obliquity. When considering the estimated pulsar inclination of ${\sim}130\degr$, this also indicates that the magnetic axis swings close to the observer's line of sight. 
Our joint polarimetric,  spectral, and timing analyses hint toward a complex accreting geometry, whereby magnetic multipoles with an asymmetric topology and gravitational light bending significantly affect the behavior of the observed source.
}


\maketitle

\section{Introduction}\label{sec:introduction}

\begin{figure}
\centering 
\includegraphics[width=0.48\textwidth]{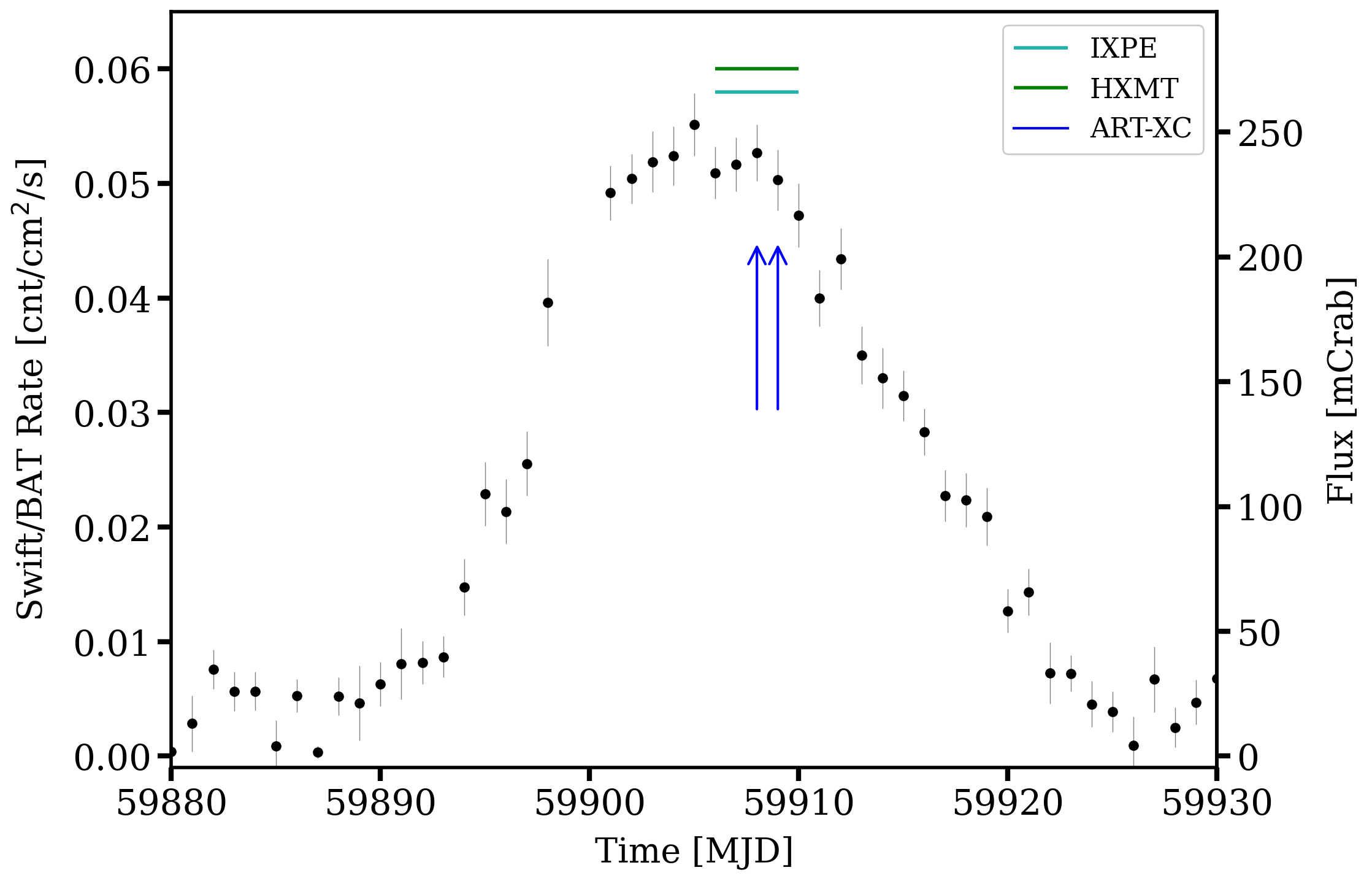}
\caption{{\it Swift}/BAT (15--50 keV) daily average light curve of \src\ (black dots with gray error bars).
Times of each continuous and pointed observations used in this work are marked by horizontal colored lines and vertical arrows, as detailed in the legend.
\label{fig:outburst}}
\end{figure}

Accreting X-ray pulsars (XRPs) are binary systems consisting of a neutron star (NS) and a donor companion star \citep[see][for a recent review]{Mushtukov22}.
In these systems, the NS can accrete matter supplied by the companion either via stellar wind or Roche-lobe overflow, thereby poducing emission in the X-ray domain.
The NS can be strongly magnetized, with a dipolar magnetic field strength on the order of $10^{12}$\,G.
This leads to highly anisotropic accretion, where the matter is funneled by the magnetic field to the magnetic poles, giving rise to pulsating X-ray emission.
Studying these systems is crucial for understanding the effects related to the interaction of X-ray radiation with strongly magnetized plasma.
In fact, the emission from XRPs can be expected to be strongly polarized, up to a polarization degree (PD) of 80\% due to magnetized plasma and vacuum birefringence \citep{Gnedin1978,Pavlov79,Meszaros1988, Caiazzo2021, Caiazzo2022}.
However, recent observations of XRPs have revealed a polarization that is far lower than expected  \citep{Doroshenko2022, Tsygankov2022,Forsblom2023}, with a phase-averaged PD of about 5\%--6\% and ranging from 5\% to 15\% in the phase-resolved analysis. 

\src\, is an XRP discovered with the {\it EXOSAT} observatory \citep{Parmar+89}, which also detected pulsations at about 42\,s.
The orbital period of 46.02 days was derived from the Type I outburst periodicity by \citet{Wilson+08}. These authors also obtained an orbital solution consisting of a rather eccentric (eccentricity of $e\sim0.41$) and wide (semi-major axis of $a_{\rm x} \sin\,i=248\pm2\,$lt-s) orbit.
Besides being the most regular and prolific Type I outburst XRP, \src\, also has shown sporadic Type II (or giant) outbursts \citep{Parmar+89,Corbet+Levine06, Thalhammer2021}.
The source spectrum showed a hint of the cyclotron resonant scattering feature (CRSF) at 36\,keV \citep{Reig+Coe98} and 63\,keV \citep{Klochkov08},  however, this has not been securely confirmed in other works.
More recently, the source spin period was measured to be around 41.2\,s \citep{Thalhammer2021}, after the source underwent a significant spin-up episode following the Type II outburst, as monitored by {\it Fermi}/GBM.\footnote{\url{https://gammaray.nsstc.nasa.gov/gbm/science/pulsars/lightcurves/exo2030.html}} 
The distance to the source is 
$2.4_{-0.4}^{+0.5}$\,kpc, as given in the {\it Gaia} Data Release 3 \citep{Bailer-Jones21}.

Here, we present the results of a multi-observatory campaign on \src. The observations by the {\it Imaging X-ray Polarimeter Explorer} (\ixpe) were supplemented by contemporaneous observations with {\it Insight-HXMT} and {\it Spectrum-Roentgen-Gamma}/ART-XC at the peak of a Type I outburst in 2022.

\section{Observations and data reduction}\label{sec:data_reduction}

\subsection{\ixpe}\label{subsec:IXPE_data_reduction}

\ixpe \citep{Weisskopf2022} is a NASA small explorer mission in collaboration with the Italian Space Agency (ASI), launched on  2021 December 9.
It features three identical Mirror Module Assembly (MMAs), each comprising of a grazing incidence telescope and a polarization-sensitive Detector Unit (DU) at its focus \citep{Baldini21, Soffitta21}.
The DUs consist of gas-pixel detectors (GPD) filled with dimethyl ether, whose interaction with X-ray photons produces photoelectrons that are ejected in a direction that is distributed as $\cos^2\varphi$, where $\varphi$ is the polarization direction of the incident radiation \citep{Bellazzini03}.
\ixpe provides imaging polarimetry over a nominal energy band of 2--8 keV, within a field of view of about 12.9 arcmin$^2$ for each MMA and with an energy-dependent polarization sensitivity expressed by a modulation factor (i.e., the amplitude of the instrumental response to $100\%$ polarized radiation) peaking at $\mu\sim$50\%--60\% at 8 keV.

\ixpe observed \src\, over the period 2022 November 23--27   (ObsID 02250201) for a total exposure of about 181\,ks.
A \textit{Swift}/BAT \citep{Gehrels_2004,Krimm13} light curve of the relevant outburst with \ixpe and other pointed observations is shown in Fig.~\ref{fig:outburst}.
\ixpe data have been reduced using the \textsc{ixpeobssim} software package \citep{Baldini22}, version 30.0.0\footnote{\url{https://github.com/lucabaldini/ixpeobssim}}, and using the CALDB version 20221020.
Source events were extracted from a 60\arcsec\  radius circle centered on the brightest pixel, while background events are negligible given the relatively high source count rate \citep{Di_Marco2023}.
The \texttt{v12} version of the weighted response files \citep{DiMarco2022} was used to produce and analyze spectral products.
Based on \citet{Silvestri2023}, we added a systematic error of 2\% to the \ixpe spectra.

\subsection{{\it SRG}/ART-XC}\label{subsec:ARTXC_data_reduction}

The Mikhail Pavlinsky ART-XC telescope \citep{2021A&A...650A..42P} carried out two consecutive observations (ObsIDs: 12210071001, 12210071002) of \src,\ from 2022 November 25-26 (MJD~59908.87--59909.62 and  59909.71--59909.83), simultaneously with \ixpe, with a total net exposure of 75~ks. 
ART-XC is a grazing incidence-focusing X-ray telescope on board the \textit{SRG} observatory \citep{2021A&A...656A.132S}. 
The telescope includes seven independent modules and provides imaging, timing, and spectroscopy in the 4--30~keV energy range, with a total effective area of $\sim 450$~cm$^2$ at 6~keV, angular resolution of 45\arcsec, energy resolution of 1.4~keV at 6~keV, and timing resolution of 23~$\mu$s.
The ART-XC data were processed with the software \textsc{artproducts} v1.0 and the CALDB version 20230228. 
We limited the ART-XC energy band to the 6.5--25\,keV energy range, where the instrument calibration is better known. 
Following standard procedures, we merged data from both observations, rebinned the spectrum to match the energy resolution of the detectors, and added a systematic error of 2\% to it.

\subsection{\it {Insight-HXMT}}\label{subsec:HXMT_data_reduction}

Hard X-ray Modulation Telescope ({\it HXMT}, also dubbed as {\it {Insight-HXMT}}) excels in its broad energy band (1--250 keV) and a large effective area in the hard X-ray energy band \citep{2020SCPMA..6349502Z}. \src\ was observed by {\it HXMT} from 2022 November 18 (MJD 59901) to November 27 (MJD 59910). In this work, we only analyze quasi-simultaneous observations with {\it IXPE} from 2022 November 23 (MJD 59905) to November 27 (MJD 59910).
The resulting total exposure times are 42\,ks, 71\,ks and 67\,ks for the detectors of three payloads on board {\it HXMT}, LE (1--15\,keV), ME (5--30\,keV), and HE (20--250\,keV), respectively.
The detectors were used to generate the events in good time intervals (GTIs). The time resolution of the HE, ME, and LE instruments are $\sim$25~$\mu$s, $\sim$280~$\mu$s, and $\sim$1 ms, respectively. Data from {\it HXMT} were considered in the range 2--70\,keV, with the exclusion of 21--24\,keV data due to the presence of an Ag feature \citep{Li2020}. Insight-HXMT Data Analysis software\footnote{\url{http://hxmtweb.ihep.ac.cn/}} (HXMTDAS) v2.05 and HXMTCALDB v2.05 are used to analyze the data. We screened events for three payloads in HXMTDAS using \texttt{legtigen}, \texttt{megtigen}, \texttt{hegtigen} tasks according to the following criteria for the selection of GTIs: (1) pointing offset angle $<$ 0.1\degr; (2) the elevation angle $>$10\degr; (3) the geomagnetic cut-off rigidity $>$8 GeV; (4) the time before and after the South Atlantic Anomaly passage $>$300 s; (5) for LE observations, pointing direction above bright Earth $>$30\degr. We selected events from the small field of views (FoVs) for LE and ME observations, and from both small and large FoVs for HE observations due to the limitation of the background calibration. The instrumental background is estimated by blocking the collimators of some detectors completely. The background model is developed by taking the correlations of the count rates between the blind and other detectors. The background is generated with \texttt{lebkgmap}, \texttt{mebkgmap}, and \texttt{hebkgmap} implemented in HXMTDAS, respectively.
We restricted the energy band for spectral analysis to 1--10, 10--30, and 30--70\,keV for LE, ME, and HE, respectively, as these ranges suffer smaller calibration uncertainties given the available observational background.
Following the official team recommendations, we added a systematic error of $1\%$ to LE and ME spectra, and $3\%$ to the HE spectrum.

\begin{figure}
\includegraphics[width=.5\textwidth]{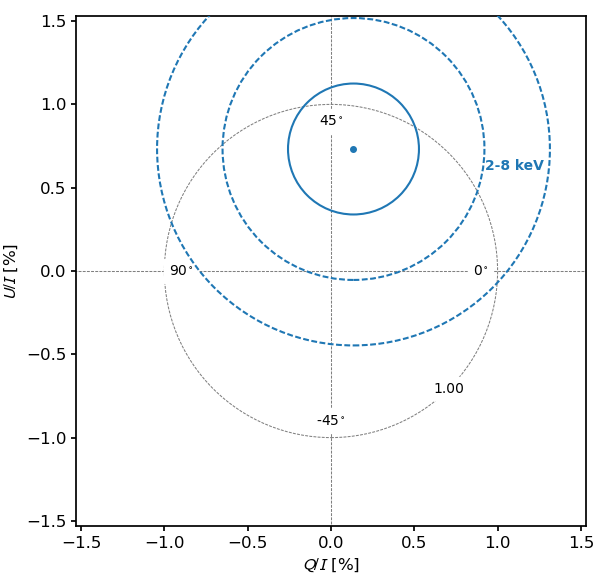}
\caption{Pulse phase-averaged normalized Stokes parameters $U/I$ (y-axis) and $Q/I$ (x-axis) over the 2--8\,keV energy range.
The $1\sigma$, $2\sigma,$ and $3\sigma$ contours are plotted as concentric circles around the nominal value (continuous and dashed lines, respectively).
The gray dotted circle represents loci of constant 1\% PD, while radial lines are labeled for specific electric vector position angles (that is, the polarization angle, PA) with respect to north.
The phase-averaged PD upper limit is about $2\%$ at $99\%$ c.l.
\label{fig:EXO2030_ETOT}}
\end{figure}

\begin{figure} 
\includegraphics[width=.49\textwidth]{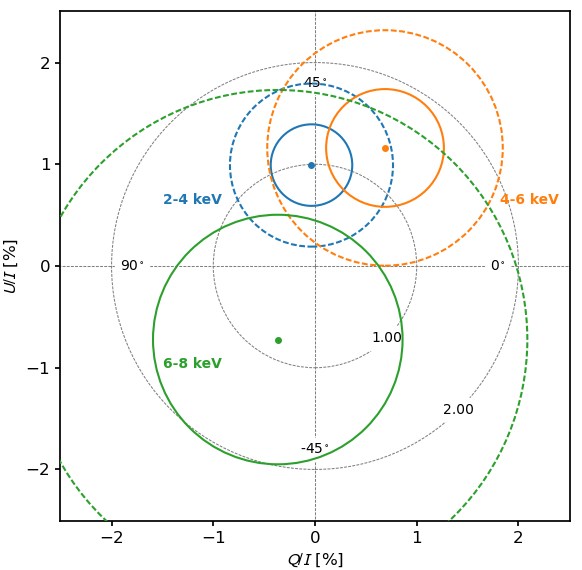}
\caption{Same details as in Fig.~\ref{fig:EXO2030_ETOT}  for energy-dependent normalized Stokes parameters.
Gray dotted circles represents loci of constant 1\% and 2\% PD.
Blue, orange, and green circles represent the 2--4, 4--6 and 6--8\,keV energy bands, respectively.
\label{fig:EXO2030_energies}}
\end{figure}

\section{Data analysis and results}

Polarimetric parameters were derived following the approach based on the formalism by \citet{Kislat2015}, as implemented in the \texttt{pcube} software algorithm and through spectro-polarimetric analysis available in \textsc{xspec} \citep{Strohmayer2017}.
The spectra were fitted simultaneously in \textsc{xspec} allowing for a cross-calibration constant to account for calibration uncertainties of different DUs with  respect to other detectors and for intrinsic source variability.
The source spectra were rebinned to have at least 30 counts per energy channel in order to adopt the $\chi^2$ fit statistic, given the non-Poissonian nature of the source spectra.
The adopted test statistic was the $\chi^2$.
Spectral data were analyzed with \textsc{xspec} version 12.13.0b \citep{Arnaud96} available with \textsc{heasoft} v6.31.

\subsection{Timing analysis}

Barycentric correction was applied to the events using the \texttt{barycorr} tool for \ixpe and \artxc, and the HXMTDAS task \texttt{hxbary} for \hxmt.
DE421 Solar system ephemeris and the SIMBAD \citep{Wenger00} ICRS coordinates of the source were employed to this aim.
Binary demodulation also was performed, employing the orbital solution from \citet{Fu2023}.
The final estimate of the spin period $P_{\rm s}$ = 41.1187(1)\,s was then obtained using the phase connection technique \citep{Deeter+81} and {\it HXMT}/LE events.
The obtained spin period was used to fold the events from all employed instruments and obtain corresponding pulse profiles.

For completeness, we also extracted the \ixpe light curve in the 2--8 keV energy band summed over the three DUs and rebinned at 300~s.
The resulting light curve shows a steady count rate of about 5~cnt~s$^{-1}$ over the whole \ixpe observation.

\begin{figure}[!ht]
\begin{overpic}[angle=-90,scale=0.35]{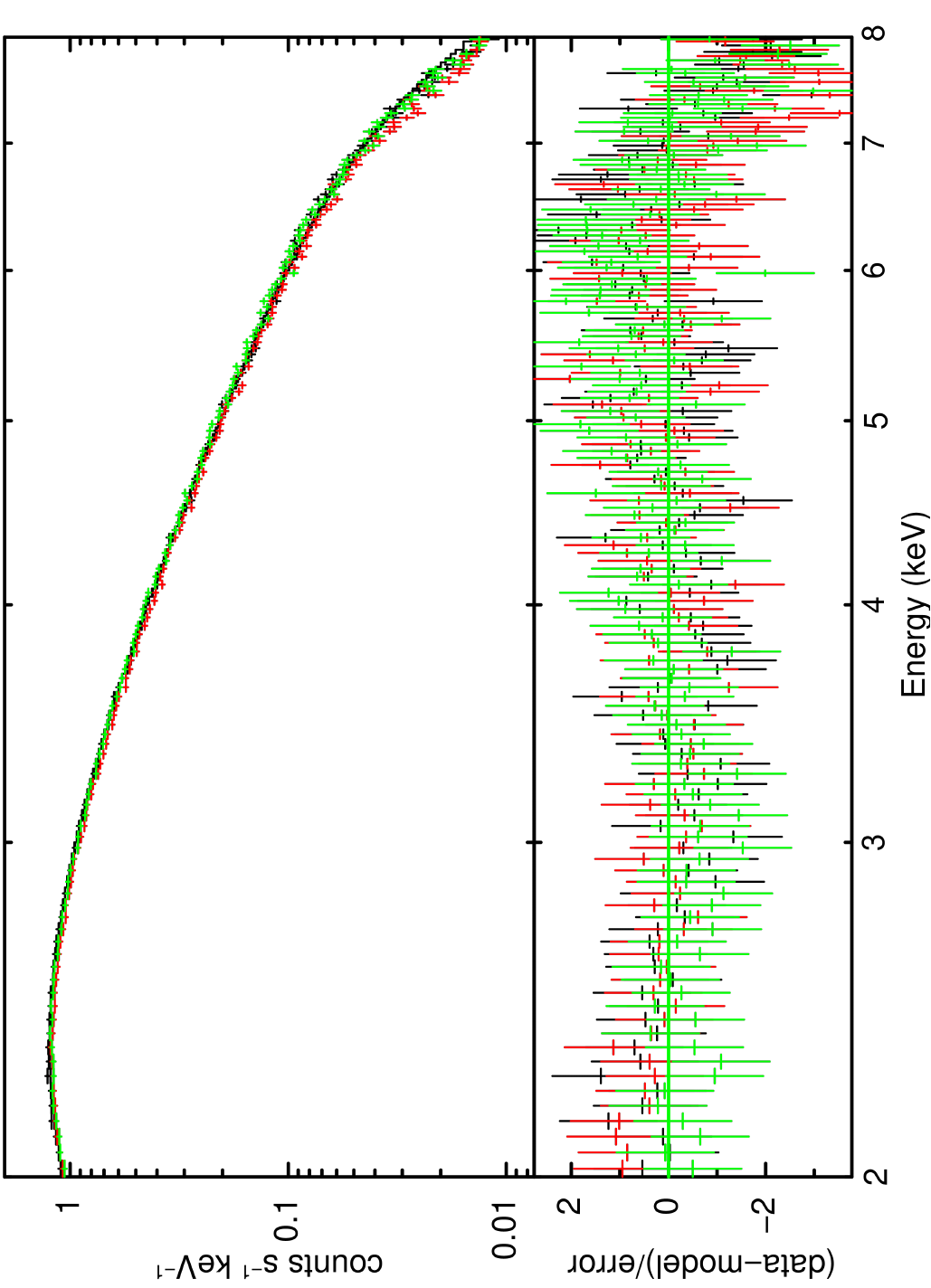}
\put(85,65){(a)}
\end{overpic}
\begin{overpic}[angle=-90,scale=0.35]{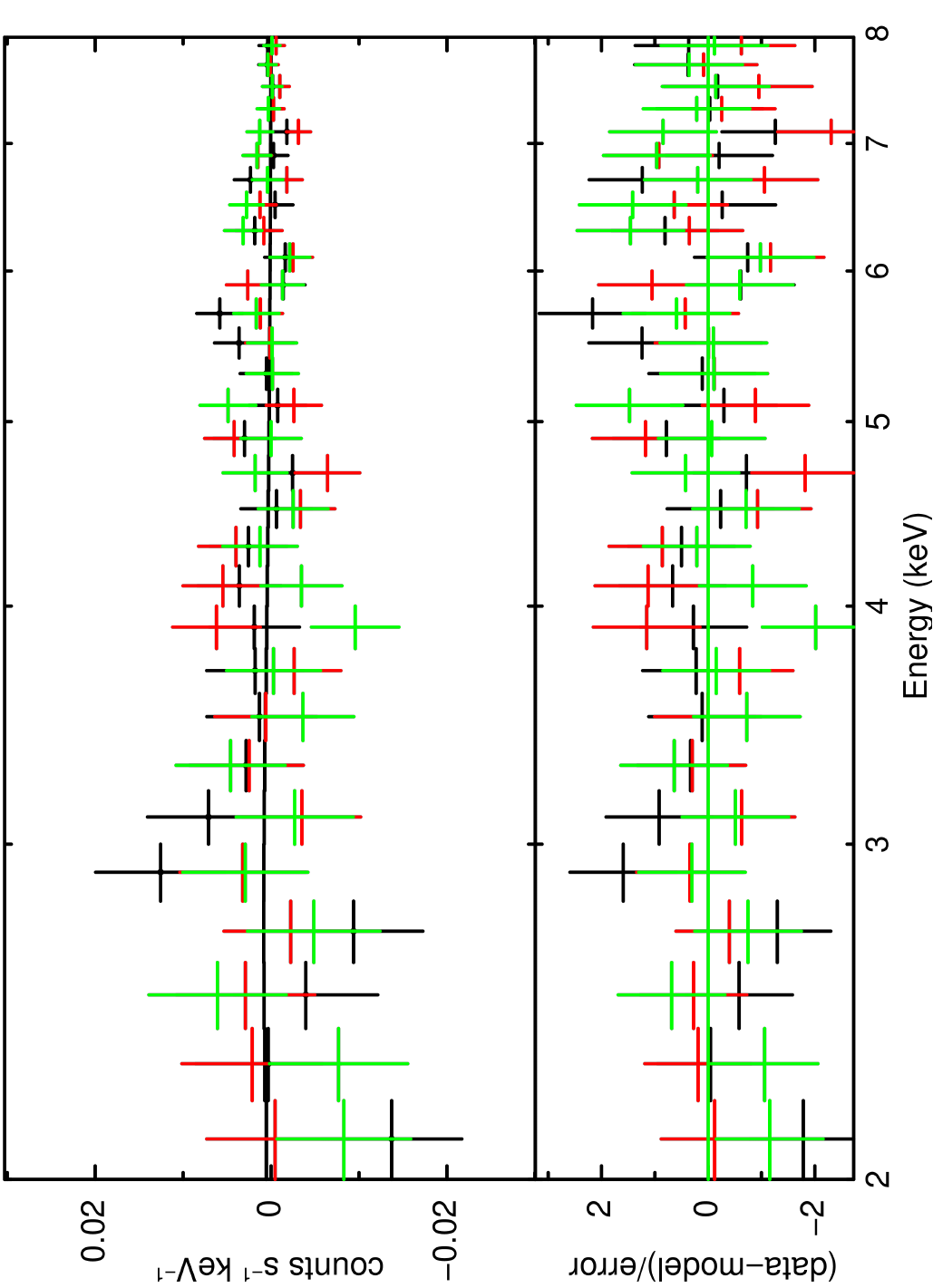}
\put(85,65){(b)}
\end{overpic}
\begin{overpic}[angle=-90,scale=0.35]{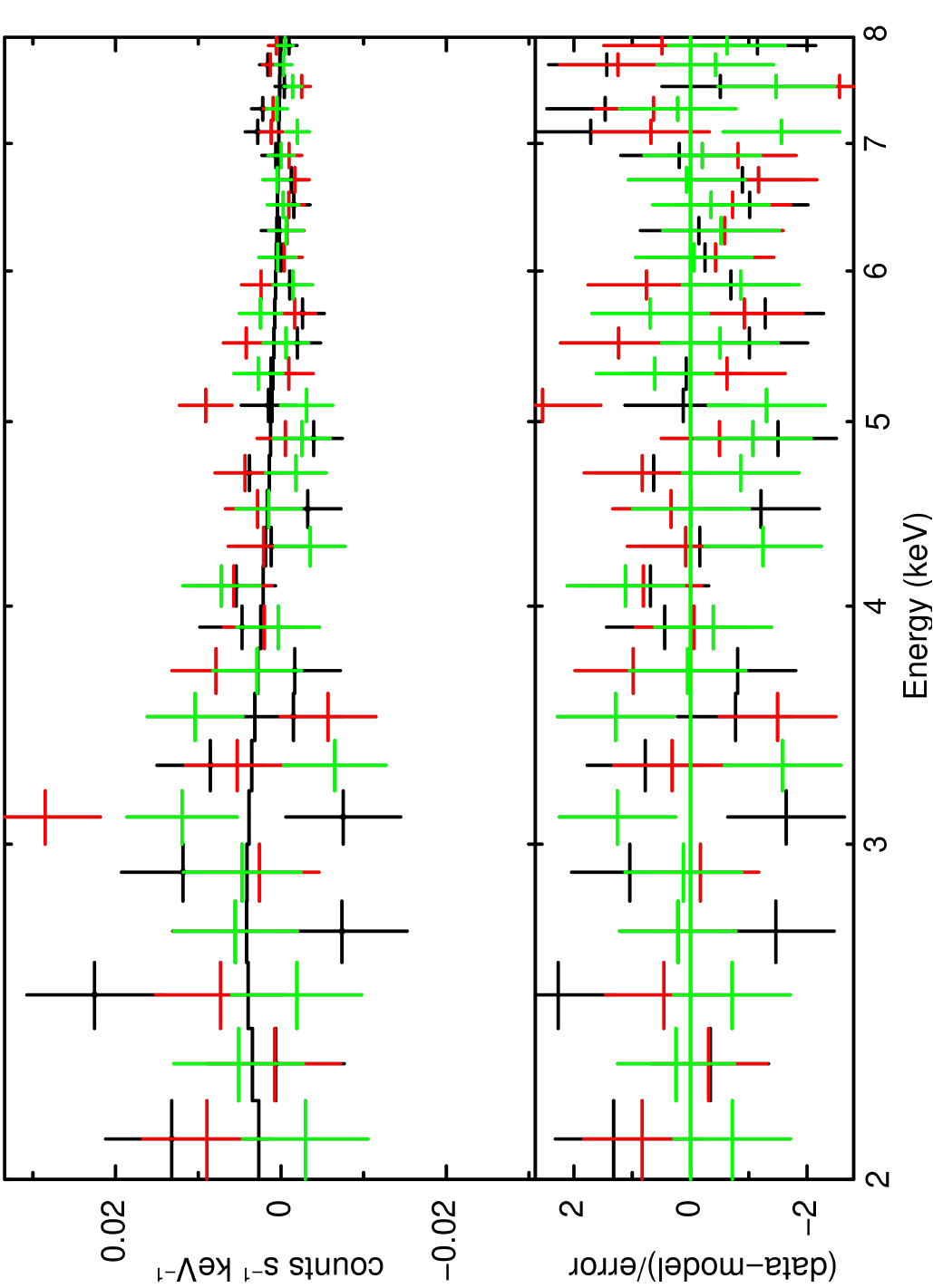}
\put(85,65){(c)}
\end{overpic}

\caption{\src\, spectral energy distribution of the phase-averaged Stokes parameters $I$, $Q$, and $U$ as observed by \ixpe\ -- panels (a), (b), and (c), respectively. Continuous lines in the top panels represent the best-fit model  \texttt{const$\times$tbabs(powerlaw$\times$polconst)} reported in Table \ref{table:phase-averaged_ixpe}. Bottom panels show the residuals. Different colors represent different detectors -- black for DU1, red for DU2, and green for DU3. Data have been rebinned and re-normalized for plotting purpose.}
\label{fig:EXO2030_spectro-polarimetric_phase-averaged_plot}
\end{figure}

\begin{table} 
\caption{Best-fit parameters of the phase-averaged \ixpe data on \src\, obtained from spectro-polarimetric analysis  using model \texttt{const$\times$tbabs(powerlaw$\times$polconst)} in the 2--8\,keV energy band. 
} \label{table:phase-averaged_ixpe}
\centering
\begin{tabular}{lc}
\hline\hline 
Parameter &  Value\\ 
 \hline 
$C_{\rm DU1}$ (fixed) & 1 \\
$C_{\rm DU2}$ & $0.963\pm0.003$ \\
$C_{\rm DU3}$ & $0.928\pm0.003$ \\
$N_{\textrm{H}}$ [$10^{22}\,$cm$^{-2}$] & $1.93\pm0.06$ \\
$\Gamma$ & $1.29\pm0.01$ \\
Norm\tablefootmark{a} & $0.315\pm0.006$ \\
PD [\%] & $1.2\pm0.4$ \\
PA [deg] & $39\pm9$ \\
Flux$_{\rm 2-10\,keV}$\tablefootmark{b} & $2.47\pm0.05$ \\
$\chi^2$/d.o.f. & $1372/1334$ \\
\hline
\end{tabular}
\tablefoot{All reported errors are at the 68\% confidence level and based on the MCMC chain values.
\tablefoottext{a}{Normalization of the power law in units of photon\,keV$^{-1}$\,cm$^{-2}$\,s$^{-1}$ at 1 keV.}
\tablefoottext{b}{Unabsorbed flux calculated for the entire model (in units of $10^{-9}\,$erg\,cm$^{-2}\,$s$^{-1}$), obtained using the \texttt{cflux} model from \textsc{xspec} as resulting from DU1.}
}
\end{table}

\subsection{Phase-averaged analysis}

\subsubsection{Phase-averaged polarimetric analysis}
\label{subsec:phase-averaged_polarimetric}

Polarization quantities were initially derived following the model-independent approach described in \citet{Kislat2015} and \citet{Baldini22}.
Normalized Stokes parameters, $Q/I$ and $U/I$, were extracted using the \texttt{pcube} algorithm within \textsc{ixpeobssim} and then used to obtain the PD and PA.
Figure~\ref{fig:EXO2030_ETOT} shows those parameters for the full 2--8\,keV energy band, while Fig.~\ref{fig:EXO2030_energies} shows the same in different energy bands.
Both plots show that the normalized Stokes parameters are consistent with zero, which implies that the PD is also consistent with zero and is lower than $\sim$3\% at $99\%$ c.l.

\subsubsection{Phase-averaged spectro-polarimetric analysis}
\label{subsec:phase-averaged_spectropolarimetric}

To perform the spectro-polarimetric analysis, we first limited the study to \ixpe data only. The $I$, $Q$, and $U$ spectra from the source were extracted using the \texttt{xpbin} algorithm for each DU.
Given the narrow energy range of \ixpe data, the spectra can be fitted with a simpler model than that required for broader-band analysis. We therefore employed a simple absorbed power-law model for the \ixpe-only analysis.
A constant polarization component (energy-independent PD and PA) was also added to the model in \textsc{xspec}.
The final form of the model was thus \texttt{const$\times$tbabs(powerlaw$\times$polconst)}.
This model returns a fit-statistic $\chi^2$/dof=1372/1334 (see Table~\ref{table:phase-averaged_ixpe} and  Fig.~\ref{fig:EXO2030_spectro-polarimetric_phase-averaged_plot}). 
Errors are calculated through MCMC simulations using the Goodman-Weare algorithm of length $2\times10^5$ with 20 walkers and $10^4$ burn-in steps.
Best-fit results are shown in Table~\ref{table:phase-averaged_ixpe}.
The analysis reveals a PD of $1.2\pm0.6\%$ at the $90\%$ c.l.  


\begin{table} 
\caption{Best-fit parameters of the phase-averaged broadband spectrum of \src\ as observed by \ixpe, HXMT and ART-XC and obtained from spectro-polarimetric analysis using the model \texttt{const$\times$tbabs(powerlaw$\times$highecut$\times$polconst+gauss)} in~the 2--70\,keV energy band.}
\label{table:total_spectrum}
\centering
\begin{tabular}{lc}
\hline\hline 
Parameter &  Value\\ 
 \hline 
$N_{\textrm{H}}$ [$10^{22}\,$cm$^{-2}$] & $1.94\pm0.03$ \\
$\Gamma$ & $1.289\pm0.006$ \\
Norm$_\Gamma$\tablefootmark{a} & $0.310\pm0.004$ \\
$E_{\rm cut}\,$[keV] & $5.8\pm0.1$ \\
$E_{\rm fold}\,$[keV] & $23.5\pm0.3$ \\
$E_{\rm K\alpha}\,$[keV] & $6.56\pm0.02$ \\
$\sigma_{\rm K\alpha}\,$[keV] & $0.24\pm0.03$ \\
norm$_{\rm K\alpha}$ [ph\,cm$^{-2}$\,s$^{-1}$] & $0.0025\pm0.0002$ \\
PD [\%] & $1.2\pm0.2$ \\
PA [deg] & $39\pm8$ \\
$C_{\rm DU1}$ (fixed) & 1 \\ 
$C_{\rm DU2}$ & $0.963\pm0.001$ \\
$C_{\rm DU3}$ & $0.928\pm0.001$ \\
$C_{\rm LE}$ & $1.400\pm0.004$ \\
$C_{\rm ME}$ & $1.343\pm0.004$ \\
$C_{\rm HE}$ & $1.255\pm0.002$ \\
$C_{\rm ART-XC}$ & $1.395\pm0.004$ \\
Flux$_{\rm 1-70\,keV}$\tablefootmark{b} & $4.44\pm0.01$ \\
$\chi^2$/d.o.f. & $2448/2592$ \\
\hline
\end{tabular}
\tablefoot{
All reported errors are at the 68\%  confidence level and based on the MCMC chain values.
\tablefoottext{a}{Normalization of the power law in units of photon\,cm$^{-2}$\,s$^{-1}$\,keV$^{-1}$ at 1 keV.}
\tablefoottext{b}{Unabsorbed flux (in units of $10^{-9}\,$erg\,cm$^{-2}$\,s$^{-1}$) calculated for the entire model, obtained using the \texttt{cflux} command from \textsc{xspec} as resulting from the \textit{HXMT}/LE data.}
}
\end{table}

\begin{table*}
\caption{
Best-fit results of the spectro-polarimetric analysis of the phase-resolved \ixpe data of \src\, using the 
\texttt{const$\times$tbabs(powerlaw$\times$polconst)} model in the 2--8 keV energy band. } \label{table:phase-resolved_ixpe}
\centering
\begin{tabular}{lcccccc}
\hline\hline 
Phase & $N_{\rm H}$ & $\Gamma$ & Norm\tablefootmark{a} & PD & PA & $\chi^2/\rm d.o.f.$  \\
  & $(10^{22}\,$cm$^{-2})$ & & & (\%) & (deg) & \\ 
  \hline 
0.00--0.18 & $2.9\pm0.1$  & $1.42\pm0.02$ & $4.9\pm0.2$ & $3.0\pm1.1$ & $45\pm9$ & $991/1037$  \\
0.18--0.26 & $3.2\pm0.2$ & $1.53\pm0.04$ & $2.3\pm0.1$ & $2.1\pm1.5$ & $81\pm21$ & $1015/1098$ \\
0.26--0.35 & $3.4\pm0.1$ & $1.44\pm0.03$ & $3.2\pm0.1$ & $6.2\pm1.3$ & $62\pm6$ & $938/974$ \\
0.35--0.44 & $2.3\pm0.1$ & $1.07\pm0.03$ & $1.4\pm0.1$ & $4.8\pm1.5$ & $38\pm11$ & $754/788$ \\
0.44--0.63 & $3.34\pm0.09$ & $1.17\pm0.02$ & $6.4\pm0.2$ & $6.3\pm0.7$ & $-6.5\pm3.3$ & $878/905$ \\
0.63--0.72 & $2.5\pm0.1$ & $1.22\pm0.03$ & $2.4\pm0.1$ & $2.9\pm1.1$ & $41\pm11$ & $1009/1048$ \\
0.72--1.00 & $3.06\pm0.07$ & $1.30\pm0.02$ & $10.9\pm0.2$ & $2.7\pm0.6$ & $89\pm5$ & $1103/1222$ \\ 
\hline
\end{tabular}
\tablefoot{All reported errors are at 68\% confidence level.
\tablefoottext{a}{Normalization of the power law in units of $10^{-2}$ photon\,keV$^{-1}$\,cm$^{-2}$\,s$^{-1}$ at 1 keV as obtained from DU1. }
}
\end{table*}

To test a possible energy-dependence of the polarization properties in \src, different polarization model components were also tested, namely \texttt{pollin} and \texttt{polpow} in \textsc{xspec}, corresponding to a linear and a power-law dependence with energy, respectively, of the PD and PA.
However, these models did not further reduce the $\chi^2$ value, nor returned significantly different polarimetric quantities and were, therefore, not explored further.

Finally, we simultaneously fitted \ixpe, HXMT, and ART-XC spectra.
In principle, with a broadband spectrum available, polarimetric results suffer less contamination from a possibly incorrect spectral model derived by the restricted \ixpe energy band.
Following previous works \citep{Klochkov08, Epili+17, Fuerst18, Tamang22}, we adopted an absorbed power-law model with high-energy cutoff and an iron K$\alpha$ line.
To this, we added a constant polarization component as above.
As the iron line is produced by fluorescence, it is not expected to be polarized. 
We verified this by adding a separate \texttt{polconst} component for the continuum and for the iron line. This resulted in a best-fit model whose PD value for the iron line was pegged at its lower limit. 
Therefore, we  left that component unaffected by polarization. 
The final model expression is thus
\texttt{const$\times$tbabs(powerlaw$\times$highecut$\times$polconst+gauss)}.
For the fitting procedure, \ixpe and {\it SRG}/ART-XC spectral parameters were tied to those from {\it HXMT}/LE, leaving a cross-calibration constant free for each instrument.
For the photoelectric absorption from neutral interstellar matter, we employed the \texttt{tbabs} model from \citet{Wilms00} and relative \texttt{wilm} abundances.
The Galactic column density in the direction of the source is about $8.8\times10^{21}\,$cm$^{-2}$ \citep{HI4PI2016}.

\begin{figure}
\centering 
\includegraphics[angle=-90,width=0.95\linewidth]{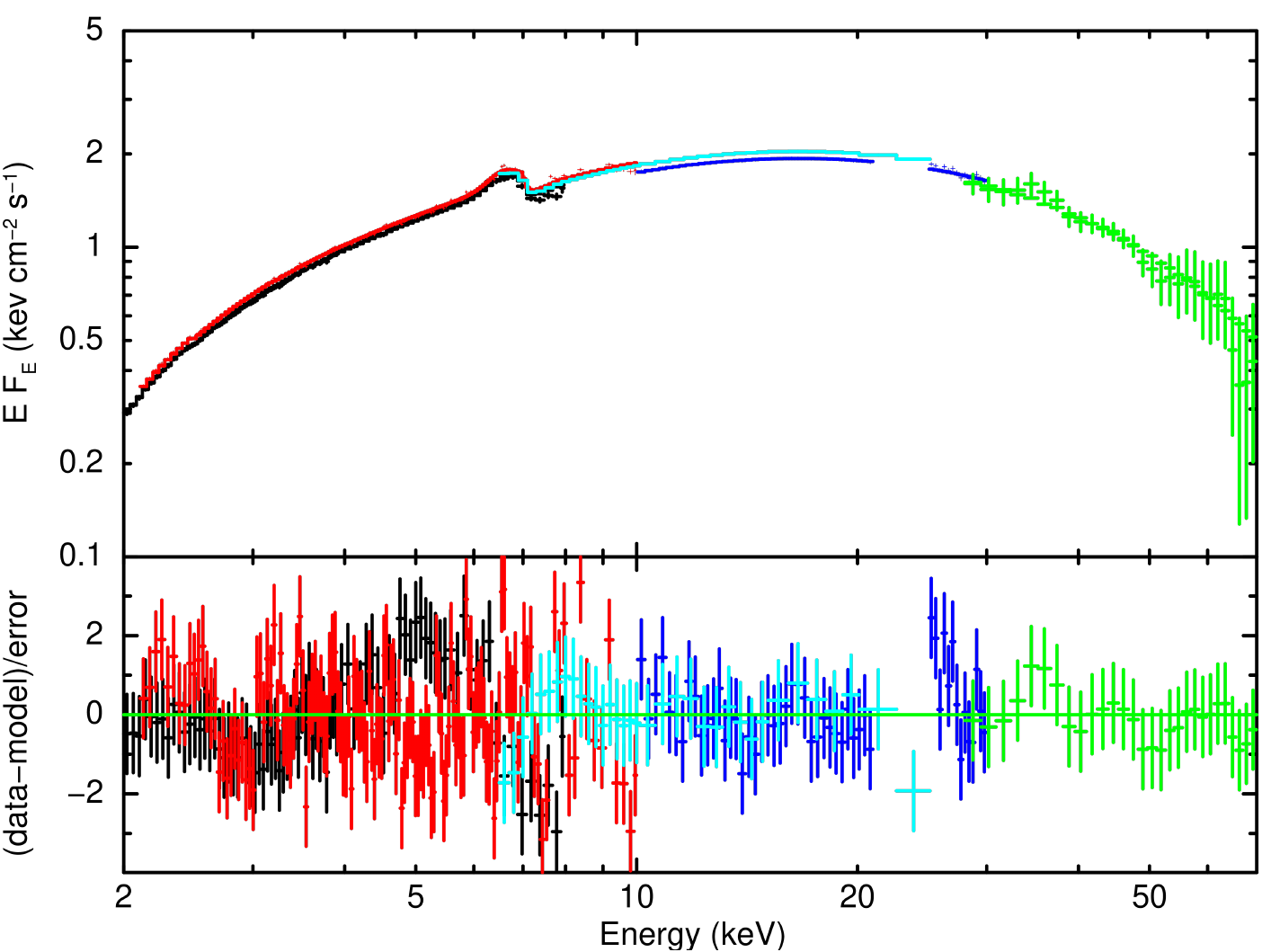}
\caption{Phase-averaged broadband spectrum of \src. \textit{Top:} \src\ unfolded $EF_E$ spectrum as observed by \ixpe, \hxmt, and \artxc. For plotting purpose, data from the three \ixpe DUs are combined and re-normalized, and all spectra are rebinned. \ixpe\ data are in red, \hxmt LE, ME, and HE in black, blue and green, respectively, \artxc data are in cyan. \textit{Bottom:} Residuals of the best-fit model (also see Table \ref{table:total_spectrum}). 
}
\label{fig:total}
\end{figure}

Despite the more elaborate model (with respect to the \ixpe-only analysis) and the broad 2--70\,keV energy band, we were still able to verify that the obtained best-fit values of the PD and PA are in agreement with those reported in Sect. \ref{subsec:phase-averaged_polarimetric} within $1\sigma$.
The broadband spectral results are shown in Fig.~\ref{fig:total} and reported in Table~\ref{table:total_spectrum}.

\subsection{Phase-resolved (spectro-)polarimetric analysis}

\begin{figure}
\centering 
\includegraphics[width=0.75\linewidth]{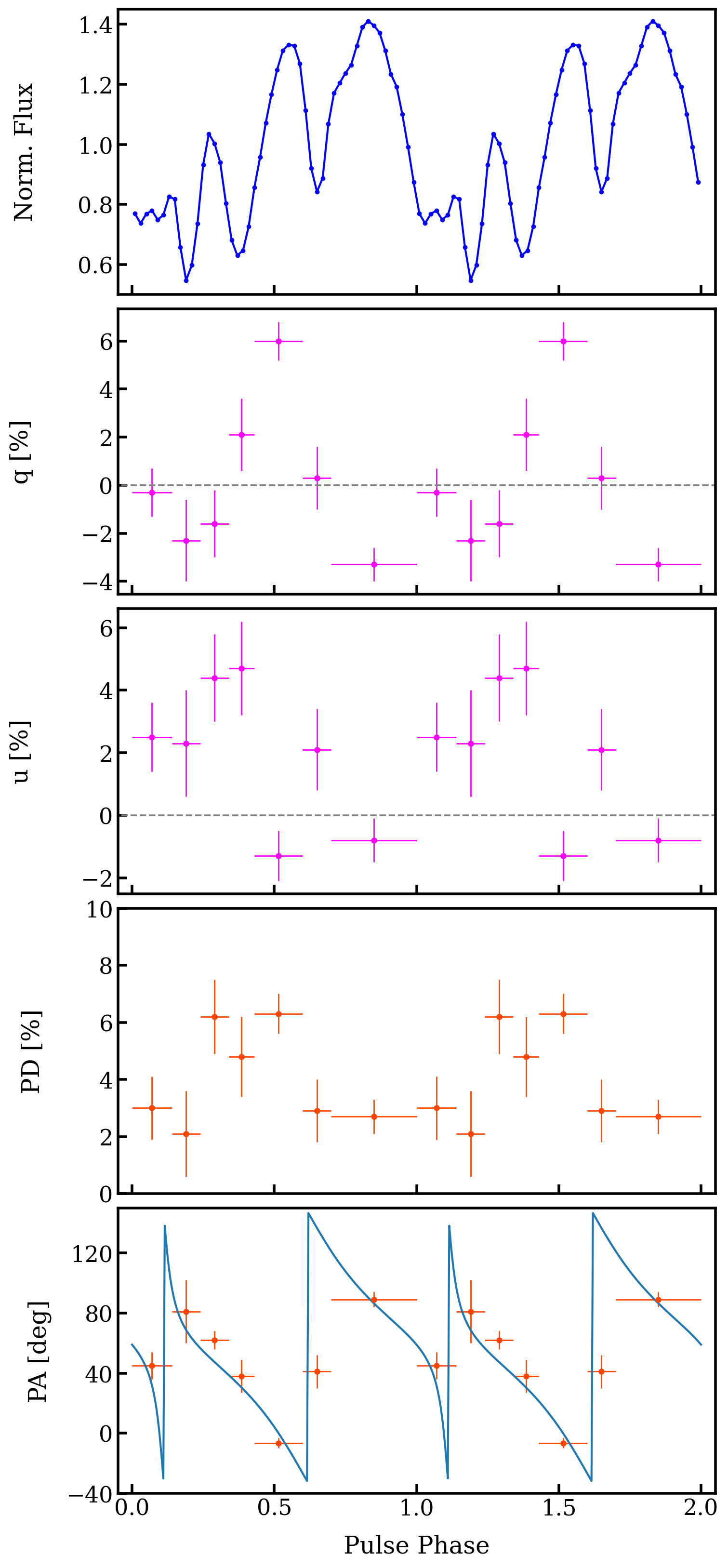}
\caption{Phase-resolved results of \src\ in the 2--8 keV range, combining data from all \ixpe DUs. From top to bottom, we show  the pulse profile, normalized Stokes parameters $q$ and $u$ based on the polarimetric analysis, and the PD and PA, as obtained from the spectro-polarimetric analysis.
The blue line in the PA panel corresponds to the best-fit rotating vector model (see Sect.\,\ref{subsec:geometry}).
}
\label{fig:phase_resolved}
\end{figure}

To perform a phase-resolved polarization analysis of \ixpe data, we selected seven phase bins to sample different flux levels shown by the pulse profile (see Fig. \ref{fig:phase_resolved}).
The phase bins were extracted with respect to $T_0=  59906.82181991$ MJD.

For the polarimetric analysis, we followed the \citet{Kislat2015} formalism as outlined in Sect. \ref{subsec:phase-averaged_polarimetric}. 
The results, shown in Fig. \ref{fig:phase_resolved},  exhibit only a moderate variability of the Stokes parameters as a function of the pulse phase. 


To perform the spectro-polarimetric analysis of the phase-resolved spectra, we used the same model as we did for the phase-averaged analysis in Sect. \ref{subsec:phase-averaged_spectropolarimetric}.
Phase-resolved spectra were rebinned analogously to the phase-averaged analysis.
For \ixpe, cross-normalization constants were kept fixed at their correspondent phase-averaged value (see Table~\ref{table:phase-averaged_ixpe}).
The resulting best-fit values are reported in Table \ref{table:phase-resolved_ixpe} and shown in Fig. \ref{fig:phase_resolved}.
The analysis reveals significant detection of polarization up to about 7\%.
Both the PD and PA show pronounced variation with spin phase.

\subsection{Phase-resolved spectral analysis}\label{subsubsec:phase-resolved_spectral}

\begin{figure}
\includegraphics[width=0.95\linewidth]{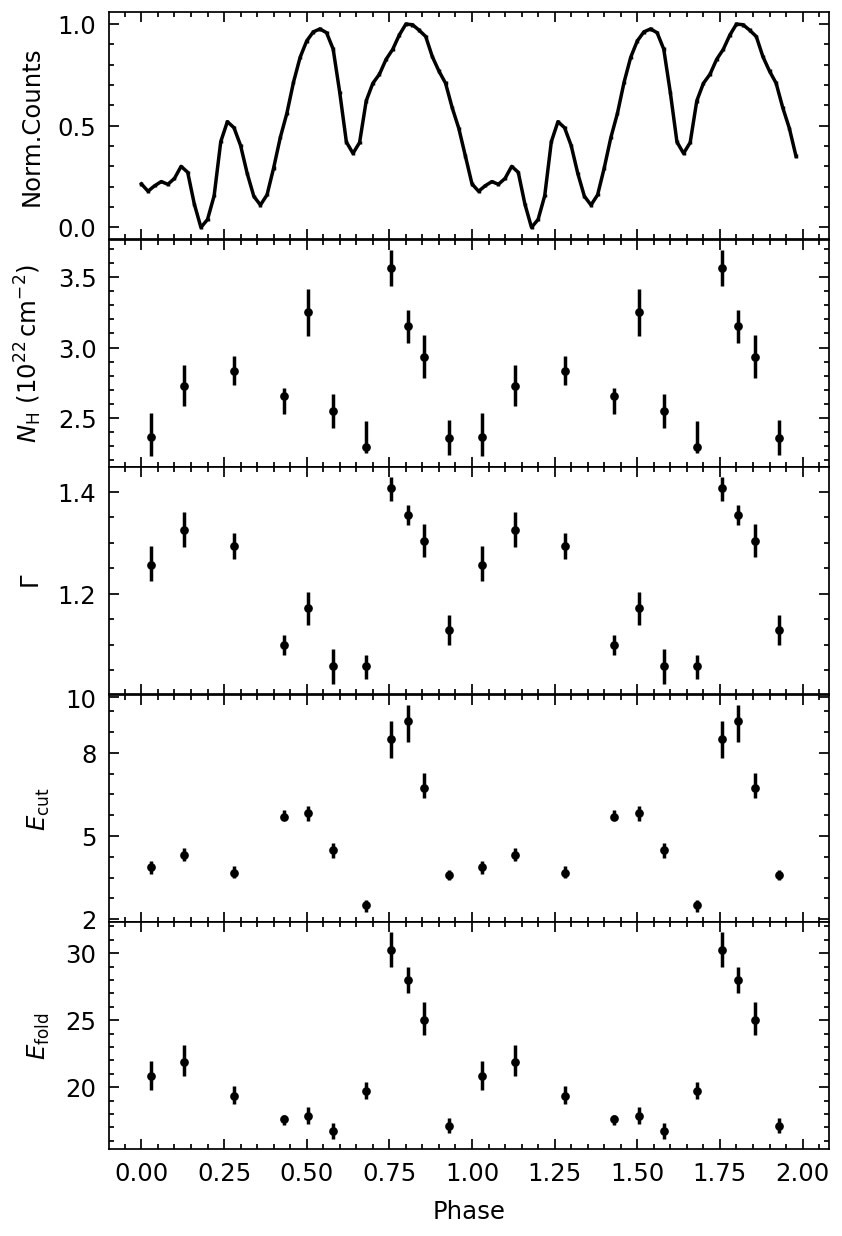}
\caption{Best-fit parameters for the broadband ($2-50\,$keV) phase-resolved spectra of \src\ as observed by {\it HXMT}. Panels from top to bottom show the pulse profile as observed by \hxmt/LE in the 2--10\,keV energy band; the column density, $N_{\rm H}$; the power-law photon index $\Gamma$; the cutoff energy; and the folding energy (both in keV).}
\label{fig:phase_resolved_HXMT}
\end{figure}

Taking advantage of the long, broadband \textit{HXMT} observations, we also performed a pulse phase-resolved spectral analysis of the \textit{HXMT} data.
For this, 11 phase bins were chosen to provide similar statistics of spectra in each bin.
However, given the limited statistics with respect to phase-averaged analysis, we limited \hxmt/HE data to 50~keV.
To model the phase-resolved spectra, we used the same model employed for the phase-averaged spectrum (see Sect. \ref{subsec:phase-averaged_spectropolarimetric}).
The best-fit results are reported in Fig. \ref{fig:phase_resolved_HXMT}.
The analysis shows strong variability of the spectral parameters with pulse phase.
We notice that the observed parameter variations might at least partly due to artificial correlations of degenerate parameters. 
We tested this through the calculation of contour plots for different pairs of parameters and verified that although the parameters show some intrinsic correlations, their variability is still significant.
Although part of this variability is known to be model-dependent \citep{Klochkov08, Hemphill2014}, it is nonetheless useful to test luminosity-dependence of the parameters variability with pulse phase (see Sect. \ref{subsec:hxmt_discussion}).

\section{Discussion}

\begin{figure*} 
\centering
\includegraphics[width=.85\textwidth]{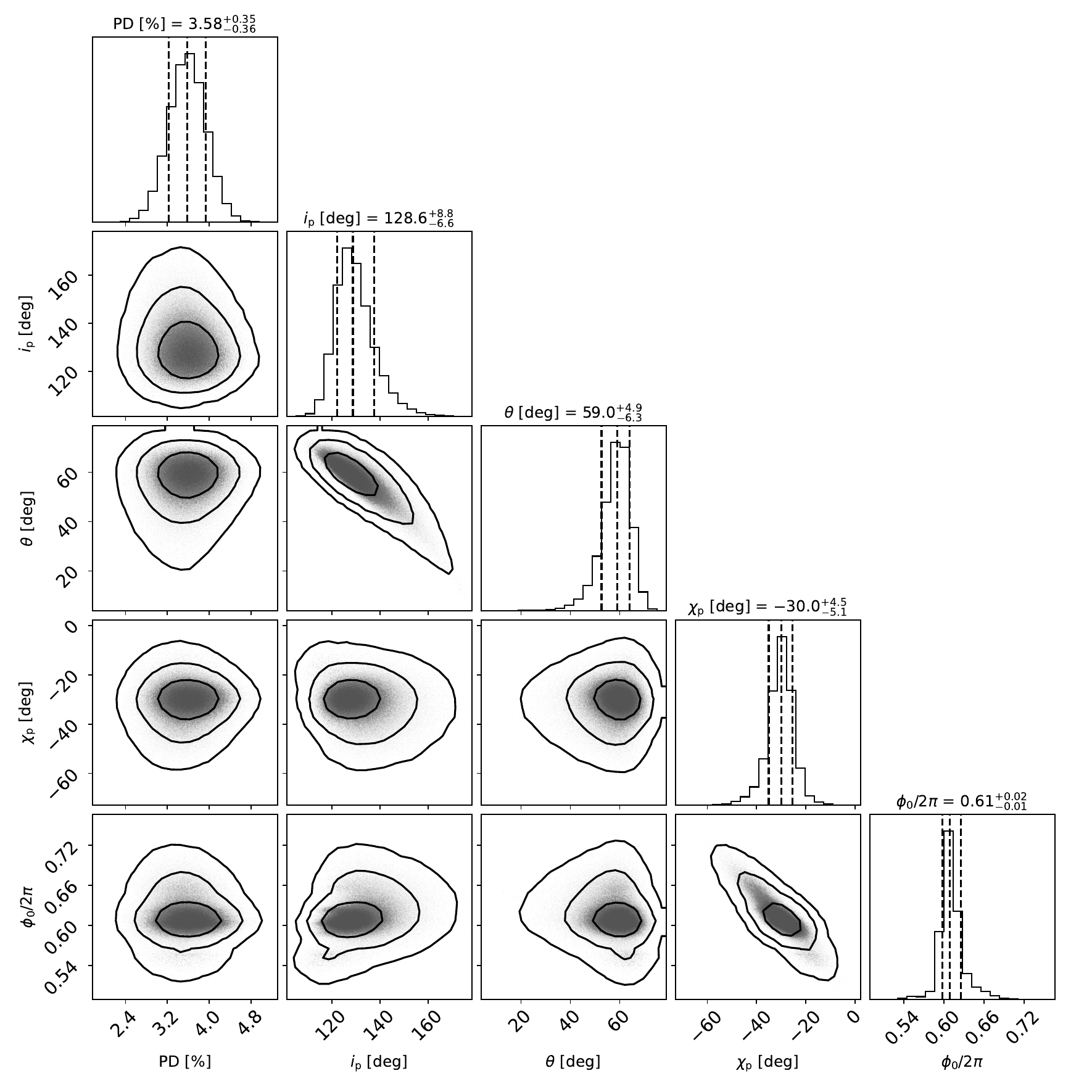}
\caption{Corner plot of the posterior distribution for the RVM parameters for the pulsar geometry obtained using the PA values. Parameters are the PD of radiation escaping from the magnetic pole, pulsar inclination, $i_{\rm p}$; magnetic obliquity, $\theta$; position angle, $\chi_{\rm p}$; and the phase, $\phi_0$.
2D contours correspond to $68\%$, $95\%$ and $99\%$ confidence levels. 
The histograms show the normalized 1D distribution for a given parameter derived from the posterior samples. The mean value and $1\sigma$ confidence levels for the derived parameters are reported above the corresponding histogram.} 
\label{fig:corner}
\end{figure*}

\subsection{Polarization degree: Expectations versus observations}

Our analysis shows a low polarization for the X-ray radiation from \src, with the phase-averaged PD in the 0\%--3\% range and the phase-resolved PD values in the range of 2\%--7\%.
High values for the PD  were expected from theoretical models of accreting XRPs \citep[and references therein]{Caiazzo2021,Caiazzo2022}.
This is due to both plasma and vacuum birefringence, which modify the opacity in the magnetic field depending on polarization of  photons.
Thus, emitted photons get polarized in two normal modes, namely: ordinary (O) and extraordinary (X), representing oscillations of the electric field parallel and perpendicular to the plane formed by the local magnetic field and the photon momentum, respectively.

Recently, however, those models have been challenged by \ixpe observations of several accreting XRPs, namely: Her X-1 \citep{Doroshenko2022}, Cen X-3 \citep{Tsygankov2022}, 4U 1626$-$67 \citep{Marshall22}, 
Vela X-1 \citep{Forsblom2023}, and GRO J1008$-$57 \citep{Tsygankov2023}. 
In fact, all those sources show a far lower PD than expected.
The observed relatively low polarization was interpreted in terms of a ``vacuum resonance'' occurring where the contributions from plasma and vacuum are equal \citep{Lai+Ho2002}.
Passing through the resonance, ordinary and extraordinary polarization modes of X-ray photons would convert to each other, with a net effect of depolarizing the radiation.
This process takes place at a plasma density $\rho_V\approx10^{-4}\,B_{12}^2\,E_{\rm keV}^2\,$g cm$^{-3}$, where $B_{12}$ is the magnetic field strength in units of $10^{12}\,$G and $E_{\rm keV}$ is the photon energy in keV.
\citet{Doroshenko2022} found that a transition layer of about 3\,g\,cm$^{-2}$ (corresponding to a Thomson optical depth of about unity) would depolarize the observed radiation consistently with the measured polarimetric quantities -- if the vacuum resonance is located in the overheated atmospheric layer, which happens in the sub-critical (or low-) accretion regime.
With the 2--10\,keV flux of $2.5\times10^{-9}\,$erg\,cm$^{-2}\,$s$^{-1}$ (see Table~\ref{table:phase-averaged_ixpe}) and at a distance of 2.4\,kpc, the observed source luminosity is $2\times10^{36}$\,erg\,s$^{-1}$.
This luminosity value is comparable to the low luminosity state of Cen X-3 \citep{Tsygankov2022} and to the bright state of GRO J1008$-$57 \citep{Tsygankov2023}, as observed by \ixpe.
The former also shows no significant polarization in the phase-averaged analysis, while the latter shows significant polarization of about $4\%$.
Therefore, it is possible that some other mechanisms beyond those linked to the accretion luminosity are responsible for the observed polarization degree. 

One qualitative interpretation of the observed low PD is pointed by the complex pulse profile of \src\ (see Figs. \ref{fig:phase_resolved} and \ref{fig:phase_resolved_HXMT}).
Such a complexity may derive from a complex magnetic field geometry where different hot spots simultaneously contribute to the observed emission at different pulse phases. 
The observed low PD might therefore be interpreted as due to mixing of emission from several parts of NS surface observed at different angles.

Another interpretation can be linked to the relation between the magnetic field geometry, in particular, the magnetic obliquity and the observer's line of sight. 
If the magnetic dipole is nearly aligned with the rotation axis and the observer looks from the side (as seems to be the case for Her X-1 and Cen X-3), the changes in the PA with the pulsar phase are rather small and the average polarization is significant. 
On the other hand, for a highly inclined dipole  (especially when observed at small inclinations), the variations of the dipole position angle (that is reflected in the PA) are large, resulting in a strongly reduced average polarization.
This interpretation is in line with the results obtained for the system geometry in \src\ and further discussed in the next section.



\subsection{Geometry of the system}
\label{subsec:geometry}

The polarimetric quantity PA can be exploited to constrain the geometry of the system by fitting the unbinned polarimetric measurements from individual photoelectric angles with the rotating-vector model (RVM, \citealt{Radhakrishnan69, Poutanen2020}). 
If radiation escapes in the O-mode, the RVM describes the PA as follows:
\begin{equation} \label{eq:pa_rvm}
\tan (\mbox{PA}\!-\!\chi_{\rm p})\!=\! \frac{-\sin \theta\ \sin (\phi-\phi_0)}
{\sin i_{\rm p} \cos \theta\!  - \! \cos i_{\rm p} \sin \theta  \cos (\phi\!-\!\phi_0) } ,
\end{equation} 
where $i_{\rm p}$ is the pulsar inclination (the angle between the pulsar spin vector and the line of sight), $\chi_{\rm p}$ is the position angle (measured from north to east) of the pulsar spin, $\theta$ is the magnetic obliquity (the angle between the magnetic dipole and the spin axis), $\phi$ is the pulse phase, and $\phi_0$ is the phase when the northern magnetic pole is closest to the observer.  The other pole makes its closest approach half a period later.
Using the RVM fit to the unbinned Stokes parameters on a photon-by-photon basis \citep{Gonzalez2023} and running Markov chain Monte Carlo (MCMC) simulations, we obtained estimates of the pulsar inclination, namely: $i_{\rm p}=129_{-7}^{+9}$~deg, 
along with the co-latitude of the magnetic pole (or magnetic obliquity), $\theta=59_{-6}^{+5}$~deg, and the position angle of the pulsar spin, $\chi_{\rm p}=\chi_{\rm p,O}=-30\pm5$~deg (see Fig.~\ref{fig:corner}).
With the pulsar inclination and magnetic obliquity angles being almost supplementary, $i_{\rm p} + \theta \approx 180\degr$, the southern magnetic pole swings close to the observer line of sight at each pulsar rotation at half a period from phase $\phi_0/(2\pi)$, that is: at $\phi=0.11_{-0.01}^{+0.02}$.  

Interestingly, in the case of \src\ the RVM suggests a relatively high magnetic obliquity.
Other XRPs (e.g., Her X-1, Cen X-3) show $\theta\approx15\degr$, while a value of $\theta\approx60\degr$ observed from \src\ is closer to the orthogonal rotator GRO J1008$-$57 \citep{Tsygankov2023}.
These results indicate that \src\ stands in between the bimodal distribution peaking at 0\degr\ and 90\degr\ of the magnetic obliquity expected for isolated NSs \citep{DallOsso2017, Lander2018}, although such results do not necessarily apply to accreting XRPs \citep{Biryukov21}.

The orbital inclination can be obtained from the orbital parameters measured by \citet{Wilson+08}. 
For the optical companion stellar mass in the range 17--20 $M_\odot$, corresponding to B0V spectral class \citep{Coe88}, and assuming a NS mass of 1.4 $M_\odot$, the inclination is in the range 49\degr--55\degr \citep[see also][]{Laplace+17}.
This value is consistent with the pulsar inclination value derived through the RVM fit because the sense of rotation cannot be determined from the X-ray pulse arrival times (i.e., solutions in the range $i_{\rm orb}=$125\degr--131\degr\ are equally probable).

\subsection{HXMT phase-resolved spectral results}\label{subsec:hxmt_discussion}

Spectral parameters are expected to show pulse phase-dependence due to the highly anisotropic accretion geometry in XRPs.
We therefore performed phase-resolved spectroscopy of \hxmt\ \src\ data (see Fig. \ref{fig:phase_resolved_HXMT}). 
Phase-resolved spectroscopy of \src\ was also performed in earlier works \citep{Klochkov08, Naik2015, Tamang22}. 
However, despite the main continuum model used in past works is similar to the one adopted here, several important differences prevent a direct comparison. 
In fact, XRP spectra are known to be luminosity-dependent \citep{Mushtukov22} and,
as a consequence, different spectral components can be adopted to fit the data collected at different luminosity levels.

For \src, the main continuum model was modified in different works with additional components such as a Gaussian absorption line around 10 keV \citep{Klochkov08} or a partial covering component \citep{Naik2015, Tamang22}.
Therefore, only a qualitative comparison can be made between the results obtained here and those from previous works. 
For example, observations carried out by \citet{Klochkov08}  of \src\ show that the power-law photon index reaches a minimum around the main pulse profile peak (corresponding to the broad main peak at $\phi{\sim}0.8$ in Fig. \ref{fig:phase_resolved_HXMT}).
A similar trend has emerged also from the phase-resolved results observed in Her X-1 \citep{Vasco2013}.
Here, in contrast, we observe a maximum value of the photon index around the same peak (see Fig. \ref{fig:phase_resolved_HXMT}).
This is likely a consequence of the luminosity difference, as the above-mentioned works derive their results in the high-luminosity accretion regime ($10^{37-38}$\,erg\,s$^{-1}$, that is near or above the critical luminosity), whereas in the present work the source has been observed at sub-critical luminosity ($\sim4\times10^{36}$\,erg\,s$^{-1}$).
Such a difference is reflected in two main aspects. 
On the one side, the accretion structure beaming pattern is expected to drastically change at different regimes.
In fact, the \src\ pulse profile observed in the high-luminosity regime (see, e.g., the 2--9\,keV panel in Fig. 2 of \citealt{Klochkov08}) exhibits substantial differences with what is observed in the present work (see top panel in Fig. \ref{fig:phase_resolved_HXMT}).
This can lead to opposite observational signatures if the observer looks through the optically deep walls of the accretion column in the super-critical regime or through the optically thin hot-spots in the sub-critical regime \citep{Mushtukov15_crit_lum, Becker2022}.
On the other hand, opposite luminosity-dependences of spectral parameters have been observed in different accretion regimes in XRPs, depending on whether a gas-mediated or a radiation-dominated shock is responsible for the infalling plasma deceleration \citep[and references therein]{Klochkov+11}.
Although such behavior has generally been observed in the pulse-averaged analysis (see, e.g., \citealt{Mueller+13,Reig+13, Malacaria+15, Diez22}), pulse-to-pulse spectroscopy hints at the possibility that similar trends are at work on shorter timescales \citep{Klochkov+11, Vybornov17, Mueller+13} and that even phase-resolved spectroscopy exhibits a pulse-phase dependence of most parameters on luminosity \citep{Lutovinov16}.

The system geometry derived in Sect. \ref{subsec:geometry} shows that the southern magnetic pole swings close to the observer line of sight at phase 0.1 (that is, half a period from $\phi_0$). 
As the observation is carried out at sub-critical accretion, an accretion column with emitting walls contributing at neighbor phases is not expected.
Thus, the main pulse profile peak at phase 0.8 is perhaps due to light bending from pencil beam emission at the magnetic poles. 
Such emission is generated in an optically thin environment at the hot-spot and it is therefore intrinsically soft, leading to a maximum of the photon index.
However, this scenario would likely produce a symmetrical behavior of the spectral parameters dependence around the phase $\phi_0$, which is not observed here.
This result, together with the highly-structured pulse profile, hints to a more complex NS configuration, such as a multipolar or asymmetric topology of the magnetic field.
This kind of magnetic field configuration has also been recently proposed for other XRPs \citep{ Postnov2013,Tsygankov2017, Israel2017,Monkkonen2022}.

\section{Summary}

Our main results can be summarized as follows:

\begin{itemize}
    \item \src\ was observed in November 2022 by \ixpe, \hxmt and \artxc at the peak of a low-luminosity Type I outburst.
    \item Only a low polarization degree of 0\%--3\% has been found in the phase-averaged analysis, while the phase-resolved analysis reveals a PD in the range of 2\%--7\%.
    \item The observed low PD can be explained in terms of an overheated NS atmosphere scenario, with additional depolarizing mechanisms possibly at work in \src. We propose that mixing of emission from several parts of the NS surface observed at different angles, on one hand, and variations of the dipole position angle resulting in changes in the PA on the other, would lead to further depolarization.
    \item By means of the rotating vector model, we constrained the geometry of the accreting pulsar. The pulsar inclination is $\sim$130\degr, almost supplementary to the magnetic obliquity angle, at $\sim$60\degr (that is, their sum is ${\sim}180\degr$).
    The obtained pulsar geometry implies that the magnetic axis swings close to the observer line of sight and the system obliquity stands between orthogonal and aligned rotators.
    \item The spectral phase-resolved analysis shows evidence that the pulse phase dependence of spectral parameters is different for different luminosities. 
    This possibly reflects changes in the accretion structure at different accretion regimes, accompanied by beam pattern changes. 
    \item  Polarimetric, spectral, and timing analyses all hint toward a complex accretion geometry, where magnetic multipoles with asymmetric topology and gravitational light bending have significant effects on the resulting spectral and timing behavior of \src.
\end{itemize}

Our analysis of \src\ characterized the X-ray polarimetric and spectral properties of the source at the sub-critical accretion regime. 
Additional future observations at different luminosities would help discerning the various mechanisms at work that shape the X-ray emission properties.

\begin{acknowledgements}
The Imaging X-ray Polarimetry Explorer (IXPE) is a joint US and Italian mission.
The US contribution is supported by the National Aeronautics and Space Administration (NASA) and led and managed by its Marshall Space Flight Center (MSFC) with industry partner Ball Aerospace (contract NNM15AA18C).
The Italian contribution is supported by the Italian Space Agency (Agenzia Spaziale Italiana, ASI) through contract ASI-OHBI-2017-12-I.0, agreements ASI-INAF-2017-12-H0 and ASI-INFN-2017.13-H0, and its Space Science Data Center (SSDC) with agreements ASI-INAF-2022-14-HH.0 and ASI-INFN 2021-43-HH.0, and by the Istituto Nazionale di Astrofisica (INAF) and the Istituto Nazionale di Fisica Nucleare (INFN) in Italy. 
This research used data products provided by the IXPE Team (MSFC, SSDC, INAF, and INFN) and distributed with additional software tools by the High-Energy Astrophysics Science Archive Research Center (HEASARC), which is a service of the Astrophysics Science Division at NASA/GSFC and the High Energy Astrophysics Division of the Smithsonian Astrophysical Observatory. 
We acknowledge extensive use of the NASA Abstract Database Service (ADS).
This research was supported by the International Space Science Institute (ISSI) in Bern, through ISSI International Team project \#495.
JH acknowledges support from the Natural Sciences and Engineering Research Council of Canada (NSERC) through a Discovery Grant, the Canadian Space Agency through the co-investigator grant program, and computational resources and services provided by Compute Canada, Advanced Research Computing at the University of British Columbia, and the SciServer science platform (www.sciserver.org). 
We also acknowledge support from the Academy of Finland grants 333112, 349144, 349373, and 349906 (SST, JP), the German Academic Exchange Service (DAAD) travel grant 57525212 (VD, VFS), the V\"ais\"al\"a Foundation (SST), the Russian Science Foundation grant 19-12-00423 (AAL, IAM, SVM, AES), the French National Centre for Scientific Research (CNRS), and the French  National Centre for Space Studies (CNES) (POP).
We thank Lingda Kong and Youli Tuo for their helpful assistance in the HXMT data analysis.
\end{acknowledgements}

\bibliographystyle{yahapj}
\bibliography{references}

\end{document}